\documentclass[aps,prl,twocolumn,reprint,superscriptaddress,english]{revtex4-1}
\usepackage[normalem]{ulem}
\usepackage[utf8]{inputenc}
\usepackage[T1]{fontenc}
\usepackage{bm}
\usepackage{graphicx}
\usepackage{float}
\usepackage{wrapfig}
\usepackage{lipsum}
\usepackage{graphicx}
\usepackage[dvipsnames]{xcolor}
\usepackage{placeins}
\usepackage[linktocpage=true,colorlinks=true,pdfborder={0 0 0},linkcolor=blue,
citecolor=blue,filecolor=yellow,urlcolor=blue,bookmarks,pdfauthor={},]{hyperref}
\usepackage{amsmath}
\usepackage{gensymb}
\usepackage{nicefrac}
\usepackage{xr}
\newcommand{\LasVegasChem}{Department of Chemistry \&{} Biochemistry, University of Nevada Las Vegas, Las Vegas, Nevada 89154, USA}
\newcommand{\NEXCL}{Nevada Extreme Conditions Laboratory, University of Nevada, Las Vegas, Las Vegas, Nevada 89154, USA}
\newcommand{\LasVegasPhys}{Department of Physics \&{} Astronomy, University of Nevada Las Vegas, Las Vegas, Nevada 89154, USA}
\newcommand{\Euro}{European X-Ray Free-Electron Laser Facility GmbH, Holzkoppel 4, 22869 Schenefeld, Germany}
\newcommand{\ETH}{Institute of Geochemistry and Petrology, ETH Z\"urich, R\"amistrasse 101, 8092 Z\"urich, Switzerland}
\newcommand{\Ham}{University of Hamburg, Institute for Experimental Physics, 22761 Hamburg, Germany}
\newcommand{\Des}{Deutsches Elektronen-Synchrotron (DESY), Notkestrasse 85, 22607 Hamburg, Germany}
\newcommand{\italy}{Department of Earth and Environmental Sciences, Universit\`a degli Studi di Milano-Bicocca, Piazza della Scienza 4, 20126, Milan, Italy}
\newcommand{\Cambridge}{Department of Materials Science \&{} Metallurgy, University of Cambridge, 27 Charles Babbage Road, Cambridge, CB3 0FS, UK}
\newcommand{\CJPJapan}{Advanced Institute for Materials Research, Tohoku University 2-1-1 Katahira, Aoba, Sendai, 980-8577, Japan}

\begin{document}

\title{Ultra-fast yttrium hydride chemistry at high pressures via non-equilibrium states induced by x-ray free electron laser}

\author{Emily~Siska}
 \email{These authors contributed equally}
 \affiliation{\NEXCL}
 \author{G.~Alexander~Smith}
 \email{These authors contributed equally}
 \affiliation{\NEXCL}
 \affiliation{\LasVegasChem}
 \author{Sergio Villa-Cortes}
    \affiliation{\NEXCL}
\author{Lewis J. Conway}
 \affiliation{\Cambridge}
 \affiliation{\CJPJapan}
  \author{Rachel J. Husband}
 \affiliation{\Des}
 \author{Joshua Van Cleave}
 \affiliation{\NEXCL}
  \affiliation{\LasVegasPhys}
\author{Sylvain Petitgirard}
 \affiliation{\ETH}
\author{Valerio Cerantola}
 \affiliation{\Euro}
 \affiliation{\italy}
   \author{Karen Appel}
 \affiliation{\Euro}
        \author{Carsten Baehtz}
 \affiliation{\Euro}
        \author{Victorien Bouffetier}
 \affiliation{\Euro}
        \author{Anand Dwiwedi}
 \affiliation{\Euro}
        \author{Sebastian G\"ode}
 \affiliation{\Euro}
     \author{Taisia Gorkhover}
 \affiliation{\Ham}
  \author{Zuzana Konopkova}
 \affiliation{\Euro}
     \author{Mohammad Hosseini}
 \affiliation{\Des}
   \author{Stephan Kuschel}
 \affiliation{\Ham}
      \author{Torsten Laurus}
 \affiliation{\Des}
        \author{Motoaki Nakatsutsumi}
 \affiliation{\Euro}
\author{Cornelius Strohm}
 \affiliation{\Des}
        \author{Jolanta Sztuk-Dambietz}
 \affiliation{\Euro}
  \author{Ulf Zastrau}
 \affiliation{\Euro}
 \author{Dean Smith}
    \affiliation{\NEXCL}
 \author{Keith V. Lawler}
 \affiliation{\NEXCL}
 \author{Chris J. Pickard}
 \affiliation{\Cambridge}
 \affiliation{\CJPJapan}
\author{Craig P. Schwartz}
 \email{craig.schwartz@unlv.edu}
 \affiliation{\NEXCL}
\author{Ashkan Salamat}
 \email{ashkan.salamat@unlv.edu}
  \affiliation{\NEXCL}
 \affiliation{\LasVegasPhys}

\begin{abstract}

Controlling the formation and stoichiometric content of desired phases of materials has become a central interest for the study of a variety of fields, notably high temperature superconductivity under extreme pressures.
The further possibility of accessing metastable states by initiating reactions by x-ray triggered mechanisms over ultra-short timescales is enabled with the development of x-ray free electron lasers (XFEL).
Utilizing the exceptionally high brilliance x-ray pulses from the EuXFEL, we report the synthesis of a previously unobserved yttrium hydride under high pressure, along with non-stoichiometric changes in hydrogen content as probed  at a repetition rate of 4.5\,MHz using time-resolved x-ray diffraction.
Exploiting non-equilibrium pathways we synthesize and characterize a hydride with yttrium cations in an \textit{A}15 structure type at 125\,GPa, predicted using crystal structure searches, with a hydrogen content between 4.0--5.75 hydrogens per cation, that is enthalpically metastable on the convex hull.
We demonstrate a tailored approach to changing hydrogen content using changes in x-ray fluence that is not accessible using conventional synthesis methods, and reveals a new paradigm in metastable chemical physics.

\end{abstract}

\maketitle

\begin{figure*}
    \includegraphics{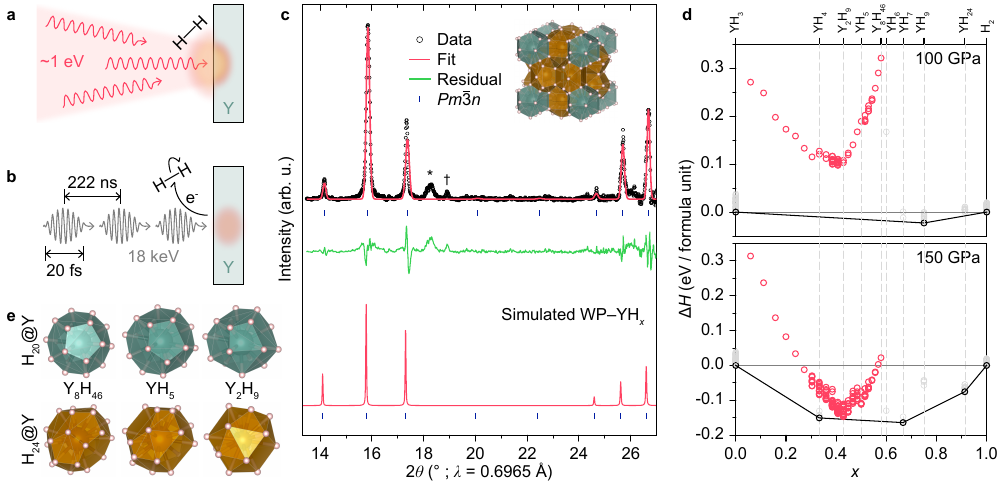}
    \caption{
    \textbf{Accessibility of non-equilibrium yttrium-hydrogen compounds \textit{via} ultrafast X-ray induced chemistry.}
    \textbf{a} Near-infrared laser heating methods primarily heat near the surface of metal samples.
    \textbf{b} Train of X-ray pulses from a free electron laser are absorbed by Y, ejecting high-energy photoelectrons which can target H--H bonds.
    \textbf{c} Le Bail refinement of YH$_{x}$ with yttrium atoms in the \textit{A}15 ($Pm\overline{3}n$) structure at 153\,GPa, (inset) structure of WP-YH$_{x}$ with ideal Y$_{8}$H$_{46}$ (i.e. Y$_{8}$H$_{46}$) stoichiometry.
    Simulated X-ray diffraction pattern of YH$_{5}$ with Y in the \textit{A}15 structure type and Weaire-Phelan-like packing of H.
    The feature marked by an asterisk (*) originates from untransformed YH$_{3}$, and the feature marked with a dagger (\dag{}) is an unknown transient feature.
    \textbf{d} Convex hull of the (H$_{3}$Y)$_{(1-x)}$(H$_{2}$)$_{x}$ system at 100 and 150 GPa, showing the increasing stability of WP-YH$_{x}$. Red circles show YH$_{x}$ species in which the Y atoms adopt an \textit{A}15-type lattice.
    \textbf{e} YH coordination polyhedra forming Weaire-Phelan volumes in YH$_{x}$.
    The top row are the cages around the corner and central metal atoms (H$_{20}$@Y in Y$_8$H$_{46}$; green) and the bottom are the cages around the face metal atoms (H$_{24}$@Y in Y$_8$H$_{46}$; orange) for Y$_8$H$_{46}$ (left), YH$_5$ (middle), and Y$_2$H$_9$ (right). The removal of hydrogen from the ideal Y$_{8}$H$_{46}$ stoichiometry reduces of the coordination of the cages and distorts the faces.
    }
    \label{fig:1}
\end{figure*}

The discovery of high-temperature superconductivity in hydrogen-dominant alloys~\cite{ashcroft_hydrogen_2004,drozdov_superconductivity_2019,einaga_crystal_2016,somayazulu_evidence_2019,snider_synthesis_2021, kong_superconductivity_2021,ma_high-temperature_2022} under high pressure has motivated the prediction and synthesis of hydrogen-rich materials spanning the entire periodic table.~\cite{bi_search_2019,flores-livas_perspective_2020,semenok_distribution_2020,hilleke_tuning_2022}
The major challenge in the synthesis of these hydrogen-rich materials has been driving ever more hydrogen atoms to coordinate with a metal center.
Hydrogen, though, is challenging to work with as it is highly diffusive and reactions can be sluggish.
Heating the metal and hydrogen reactants can improve reaction rates at high pressures,~\cite{drozdov_superconductivity_2019} and is typically achieved by focused near-infrared (Nd:YAG or similar) lasers.
Laser heating in this fashion preferentially heats near the surface of the metal through direct absorption, in a similar manner to what would be achieved by heating thermally to an equilibrium state (Figure\,\ref{fig:1}a).
Ultimately, this heating method has limited success due to the difficulty in dissociating molecular hydrogen and the sluggishness of hydrogen diffusion into the metal but has the advantage of laser-based heating being readily accessible.
Many of the interesting proposed hydrides have yet to be synthesized; the prime example being the $Fm\overline{3}m$ phase of YH$_{10}$ which has been predicted to be dynamically stable with a high-$T_{c}$ at 300\,GPa~\cite{heil_superconductivity_2019} and close to the convex hull even at high temperatures at such pressures.~\cite{liu_potential_2017,peng_hydrogen_2017,troyan_anomalous_2021}

New, less conventional methods for inducing reactions are emerging which are far from equilibrium. 
In particular, synthesis via x-ray irradiation shows great promise for allowing the synthesis of materials which otherwise could not have formed.~\cite{puglisi_X-ray_2018, stumpf_role_2016}
When high energy photons interact with matter, electrons and even atoms with very high velocities can be produced, creating unique energetic pathways that can lead to the formation of different products (Figure\,\ref{fig:1}b). 
One example of this mechanism is exciting a nitrogen atom in a NO$_{3}$ group which leads to state with an extremely short lifetime and efficient transfer of energy to the oxygen atoms which leave with higher velocities than are seen with lower energy radiation.~\cite{vinson_quasiparticle_2016,vinson_resonant_2019}
This rapid transfer of energies to atoms participating in bond breaking has also been seen for hydrogen atoms.
Water undergoes the rapid ejection of a hydrogen atom following excitation of the oxygen atoms.~\cite{takahashi_interpretation_2022, tokushima_high_2008, fransson_X-ray_2016}
At these high kinetic energies, hydrogen should be able to diffuse more rapidly than with lower energy based heating, allowing for the possibility of novel reactions.
In the case of hydrogen and superhydrides, the primary and secondary electrons from the x-ray excitation may also provide an avenue for facilitating the dissociation of molecular H$_{2}$ (Figure~\ref{fig:1}b), thus leading to increased reaction rates.

One of the biggest factors in enabling x-ray photochemistry studies is the advent of high-flux, high repetition rate free electron lasers.~\cite{cerantola_new_2021}
These instruments provide several orders of magnitude more intense pulses, which are several orders of magnitude shorter in pulse length, than synchrotron sources.
This increased temporal and spatial resolution can be utilized to carefully control the atomic state of materials and drive reactions.~\cite{eichmann_photon-recoil_2020, oneal_electronic_2020}
In the case of metal superhydrides, the heavy element can be specifically targeted energetically through its unique electronic structure.
This high x-ray flux synthetic method has previously been used in the x-ray ``heating'' of iron in the presence of nitrogen to drive a reaction under high-pressure high-flux conditions.~\cite{hwang_X-ray_2021}   
In this study, samples of yttrium in a bath of the excess reagent hydrogen, confined in a diamond anvil cell, were exposed to high fluence x-rays; producing yttrium polyhydrides, specifically a metastable \textit{A}15 structure type that has not been accessed in the yttrium hydride system using other techniques.

\begin{figure}[ht!]
    \includegraphics{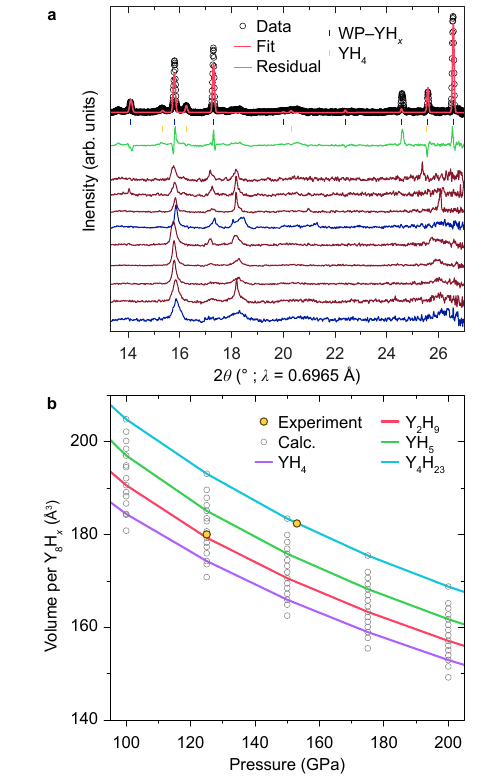}
    \caption{
    \textbf{Emergence of WP-YH$_{x}$ and its stability.}
    \textbf{a} Bottom to top: Stack plot of phase evolution from YH$_3$ as a function of time. Blue patterns represent the first XRD taken in a train. Pink patterns represent consecutive XRD measurements within the train. Not all XRD patterns collected within a train are shown for simplicity. No major changes to the sample occurred during XRD measurements not shown. 
    Above stack plot is the first XRD pattern taken for a subsequent train with Rietveld refinement of $Pm3n$ WP-YH$_x$ and $I4/mmm$ YH$_4$ shown in red.
    \textbf{b} Volumes of experimentally synthesised WP-YH$_{x}$ at 125 and 153\,GPa alongside calculated pressure-volume relations of WP-YH$_{x}$ phases. 15 WP-YH$_{x}$ stoichiometries were modeled (grey circles), and we highlight $x$\,=\,4, $x$\,=\,4.5 which sits at the minimum of the \textit{A}15 hull (Figure~\ref{fig:1}d and Figure~S7), $x$\,=\,5 which has been predicted for Eu~\cite{semenok_novel_2021}, and the saturated $x$\,=\,5.75.
    }
    \label{fig:2}
\end{figure}

\subsection{Synthesis of Weaire-Phelan type yttrium hydrides}

For these experiments, three different diamond anvil cells containing $Fm\overline{3}m$ YH$_3$ with an H$_2$ pressure transmitting medium (and excess reagant) were prepared at pressures of 104, 125 and 153\,GPa as described in the methods section. 
While these samples were all initially loaded as a Y foil, at no time during any of the experiments was the presence of Y metal in the $C2/m$ structure nor the predicted $Fddd$ high pressure structure observed,~\cite{samudrala_structural_2012, li_new_2019} either from the starting elements or possible decomposition by-products. 
Following XFEL irradiation, a new phase was seen to emerge at 125 and 153\,GPa.
This diffraction pattern of the new phase cannot be indexed by the $Fm\overline{3}m$ YH$_{1}$, $Fm\overline{3}m$ YH$_{2}$, $Fm\overline{3}m$ YH$_{3}$, $I4/mmm$ YH$_{4}$, $Im\overline{3}m$ YH$_{6}$, $P1$ YH$_{7}$, $P6_{3}/mmc$ YH$_{9}$, $P\overline{1}$ YH$_{9}$, $F4\overline{3}m$ YH$_{9}$, or $C2/c$ YH$_{12}$ structures previously reported to be on the convex hull at 150\,GPa and various temperatures.~\cite{liu_potential_2017,peng_hydrogen_2017,troyan_anomalous_2021}
Rather, this new phase is better fit by a primitive cubic structure.

To explore other possibilities along the convex hull, we trained an Ephemeral Data Derived Potential (EDDP) on the Y-H system and used it to perform extensive ab initio random structure searching (AIRSS) searches on systems with stoichiometries Y$_{2-8}$H$_{4-80}$.~\cite{pickard_high-pressure_2006, pickard_ab_2011}  
The best structures were then re-optimised using \textsc{castep}.~\cite{clark_first_2005} 
In the search, we recovered the known Weaire-Phelan (WP) $Pm\overline{3}n$ $M_8$H$_{46}$ (also denoted as $M_4$H$_{23}$ or $M$H$_{5.75}$) structure type~\cite{weaire_counter-example_1994,semenok_novel_2021,pena-alvarez_synthesis_2021} and several similar structures with hydrogen vacancies.
As shown in Figure~\ref{fig:1}c, the WP-type structure provides a good fit for this new phase.
Noting that the more hydrogen deficient structures were closer to the convex hull, we then explicitly generated hydrogen-vacancy structures starting from the fully occupied $Pm\overline{3}n$ Y$_8$H$_{46}$ (Fig~\ref{fig:1}c inset).
Figure~\ref{fig:1}d shows the DFT-relaxed convex hull at 100 and 150\,GPa where the lowest energy WP-type structure is Y$_2$H$_{9}$.
Given the many metastable stoichiometries that adopt the WP structure type with similar diffraction patterns, it is difficult to assign an exact stoichiometry and we thus describe the new phase as WP-YH$_x$ with $4\,\leq x\,\leq\,5.75$.

A Le Bail refinement of WP-YH$_x$ at 153\,GPa yields $a$\,=\,5.647(1)\,\AA{} and $V$\,=\,180.1(3)\,\AA$^3$ (Figure\,\ref{fig:1}c).
The WP-YH$_{x}$ phase observed at 125\,GPa is accompanied by the presence of $I4/mmm$ YH$_4$. 
A Rietveld refinement of a representative XRD pattern of WP-YH$_{x}$ and YH$_4$ at 125\,GPa may be seen in Figure~\ref{fig:2}a along with a stack plot illustrating the phase evolution via pulse trains. 
The lattice parameters at 125\,GPa are $a$\,=\,5.671(1)\,\AA{} for WP-YH$_x$, and $a$\,=\,2.795(1)\,\AA{} and $c$\,=\,5.218(1)\,\AA{} for $I4/mmm$ YH$_4$.
The synthesis of WP-YH$_x$ and YH$_4$ is consistent, as they are formed multiple times across the two different pressures.
Also, 125\,GPa is the lowest pressure that $I4/mmm$ YH$_4$ has been synthesized.

While not predicted for the yttrium--hydrogen system, the WP-type $Pm\overline{3}n$ $M_8$H$_{46}$ structure has been observed for both Eu and Ba.~\cite{semenok_novel_2021,pena-alvarez_synthesis_2021} 
The Weaire-Phelan structure is the current leading solution on how to partition three dimensional space into equally sized packed volumes with minimal surface area, and is observed in some clathrates.~\cite{weaire_counter-example_1994}
The metal sublattice of the WP-type structure has the $\beta$-W (ie. \textit{A}15) structure,~\cite{hartmann_elektrolysen_1931,frank_complex_1959} and the hydrogens in M$_8$H$_{46}$ are the vertices of the equally sized packed volumes.
It should be noted that the refined volumes for the $Pm\overline{3}n$ phases here are too large to be explained by the formation of an \textit{A}15 phase of pure yttrium.
In the ideal WP-type structure, the corner and central metal atoms have H$_{20}$@M cages made entirely of pentagonal faces, and the face metal atoms have H$_{24}$@M tetrakaidecahedron cages made of hexagonal and pentagonal faces.
As hydrogen vacancies are introduced, these polyhedra begin to lose vertices and distort as is illustrated for the progression from Y$_8$H$_{46}$ to YH$_5$ to YH$_{4.5}$ in Figure~\ref{fig:1}e.
A consequence is that with vacancies there are now interstitial void spaces in the polyhedral packing, thus deviating from the Weaire-Phelan packing.

Figure~\ref{fig:2}b shows the computed equations of state (EOS) curves for some of the predicted WP-type vacancy structures.
From this one can see that introduction of hydrogen vacancies does reduce the volume of the structure for a given pressure.
Likewise, the removal of hydrogen does decrease the compressibility of the WP-type hydride.
When plotted, the two experimental WP-YH$_x$ refinements show a nearly flat pressure-volume relationship, in stark contrast to the computed EOS curves.
From this it can be inferred that the WP-YH$_x$ made at 153\,GPa has a higher hydrogen content than the one made at 125\,GPa.
However, even with this comparison we are remiss to assign an exact stoichiometry owing to the uncertainties in the experimental pressures and the theoretical volumes (from possible functional, thermal and nuclear effects).



It is known that pressure alone is enough to drive the rare earth (RE) metals yttrium and lanthanum to react with hydrogen over time, yielding MH$_x$ with $0\,\geq x\,\geq 3$ using only mild compression and $x = 4, 6$ at pressures between 200\,--\,244\,GPa.~\cite{huiberts_yttrium_1996,kume_high-pressure_2007, machida_long-period_2007, palasyuk_hexagonal_2005, machida_x-ray_2006,kong_superconductivity_2021,snider_synthesis_2021,troyan_anomalous_2021, purans_local_2021}
However, to achieve higher hydrogen stoichiometry, heat has been traditionally employed to induce such reactions.~\cite{troyan_anomalous_2021, kong_superconductivity_2021, snider_synthesis_2021}
In the case of yttrium, previous studies that relied on laser heating to produce temperatures in excess of 1500\,K to induce reactions were able to synthesize higher stoichiometric hydrides such as YH$_6$ and YH$_9$, but only above 160\,GPa.~\cite{troyan_anomalous_2021,kong_superconductivity_2021}
However, there was no evidence of the $Im\overline{3}m$ phase of YH$_6$ despite those previous studies seeing it at pressures as low as 147\,GPa,~\cite{troyan_anomalous_2021,kong_superconductivity_2021} and it being predicted to be stable as low as 72\,GPa.~\cite{peng_hydrogen_2017, heil_superconductivity_2019, liu_potential_2017, li_pressure-stabilized_2015}
Likewise, no hexagonal phase of YH$_9$ was observed at 153\,GPa here despite being predicted to be on the convex hull at 150\,GPa.~\cite{peng_hydrogen_2017} 
Based on the conditions known for YH$_6$ formation, we ascribe its (and hcp-YH$_9$'s) non-formation here as a consequence of the non-equilibrium, non-thermal pathways along which reactions were induced. 
In addition, it has been shown that ultrafast dynamic compression can kinetically hinder certain phases.~\cite{smith_ultrafast_2008}

\subsection{Effects of Ultrafast Energy Deposition}

Ultrafast deposition of high energy density creates hot electrons on time scales much faster than thermalization of the lattice, leading to non-equilibrium states that can not be accessed under normal thermodynamic conditions.
Hot electrons can relax through collisional pathways or a number of different electronic processes that can lead to phase transitions in metals and semiconductors.~\cite{medvedev_nonthermal_2020,medvedev_femtosecond_2019}
It has even been predicted that high energy density could induce known phase changes at lower pressures, such as the A7 to simple cubic transition in As; observed at 26\,GPa under normal thermodynamic conditions, occurring at 23\,GPa when exposed to high enough energy density.~\cite{zijlstra_laser-induced_2008} 
Analysis of the XRD patterns taken during irradiation could provide insight to this point.
In this study, during irradiation, the (200) peak of YH$_3$ grows in intensity, possibly relating to phase fraction changes, but disappear after irradiation. On the other hand, certain peaks associated with WP-YH$_x$ and YH$_4$ decrease in intensity or completely disappear during a pulse train --- but always reappear afterward.
This phenomenon can be seen in Figure~S5 and suggests that synthesis and growth of these phases seem to be interconnected or driven by similar phenomena. 

\begin{figure*}
    \includegraphics{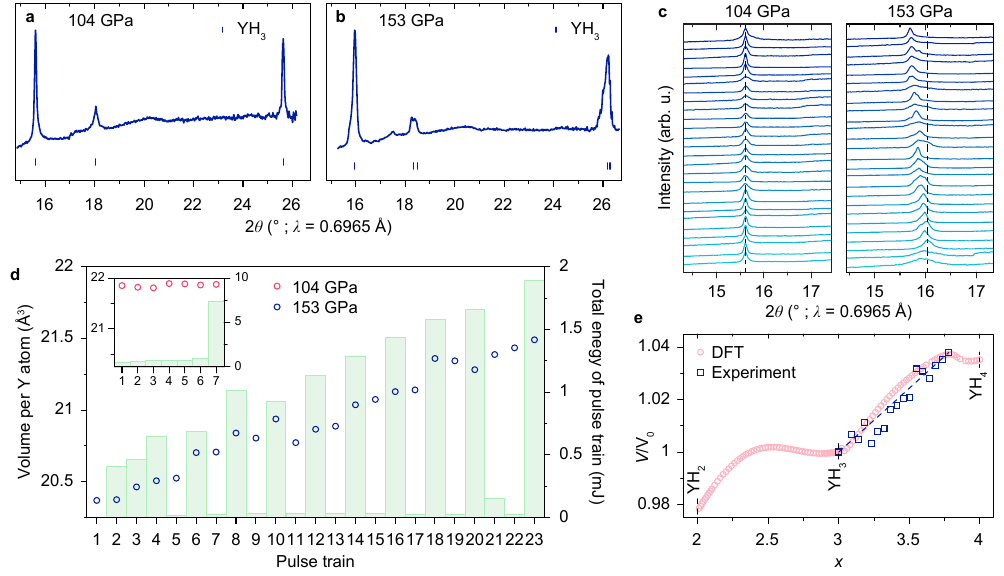}
    \caption{
    \textbf{Lattice changes in YH$_{3}$ following XFEL irradiation.}
    \textbf{a} XRD pattern showing that, at lower pressures, YH$_{3}$ remains cubic.
    \textbf{b} XRD pattern showing that, at higher pressures, X-ray irradiation causes a tetragonal distortion evidenced by splitting of the (200) and (211) peaks. Here $I4/mmm$ YH$_3$ has a $c/a$\,=\,1.43.
    \textbf{c} ``Cold'' XRD patterns from sequential pulse trains at 104 and 153\,GPa show markedly different response of YH$_{3}$ to X-ray irradiation, with the lower pressure hydride returning to its lattice after each train, and higher pressure samples exhibiting permanent swelling of the unit cell alongside the tetragonal distortion.
    \textbf{d} YH$_{3}$ at 153\,GPa shows lattice expansion which is linked to pulse train energy, more intense trains (green bars) are accompanied by larger volume changes (inset) The same effect is not seen at lower pressures, with YH$_{3}$ unit cell volumes staying near constant even following pulse trains with 7.4\,mJ total energy.
    \textbf{e} Relative X-ray driven volume changes correlate with volume expansion along a YH$_{3}\rightarrow{}$\,YH$_{4}$ transformation, measured volume changes at 153\,GPa (blue squares) co-align with simulated volumes in YH$_{x}$.
    }
    \label{fig:3}
\end{figure*}

The study by Pace et al., has suggested that the total dose of radiation is a critical parameter for photochemical reaction,~\cite{pace_intense_2020} and we have investigated this parameter within our systems. 
When the sample of YH$_3$ at 104\,GPa was irradiated, no change in the sample was observed. 
The sample was exposed with up to 254\,$\mu$J/pulse and with trains containing up to 200 pulses. 
Seemingly sufficient fluence incident on the sample with no change indicated that the cubic YH$_3$ phase is the overwhelmingly thermodynamically preferred state at 104\,GPa.
The YH$_3$ starting material was confirmed by indexing all observed peaks with a cubic lattice in the $Fm\overline{3}m$ space group.
The material had a lattice parameter $a$\,=\,4.436(5)\AA{} as refined from the patterns like those shown in Figure~\ref{fig:3}a, consistent with previous studies.~\cite{purans_local_2021, kong_superconductivity_2021} 

\begin{figure}
    \centering
    \includegraphics{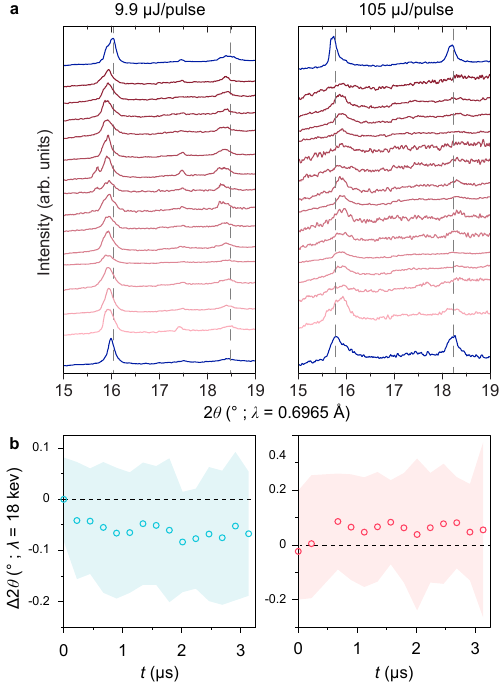}
    \caption{
    \textbf{Behaviour of YH$_{3}$ under XFEL radiation dependent on fluence.}
    \textbf{a} Evolution of X-ray diffraction signatures from samples irradiated with low (9.9\,$\mu{}$J/pulse) and high (105\,$\mu{}$J/pulse) fluence pulse trains at 153\,GPa. Dashed lines show starting positions of (111) and (200) Bragg peaks during initial ``cold'' shots.
    \textbf{b} Shift -- $\Delta{}2\theta{}$ -- in centroid of (111) Bragg peak in low and high fluence pulse trains. Shaded regions denote FWHM of peaks. Low fluence pulse trains have the effect of expanding the lattice, while high fluence trains contract the lattice.
    }
    \label{fig:4}
\end{figure}

The sample at 153\,GPa was irradiated at 3 different locations, and each location went through its series of trains with increasing fluence before moving onto the next location.
The first location had an irradiation regime and outcome distinct from the other two, most likely attributable to the starting fluence incident on the sample and the slow rate at which it was increased compared to locations 2 and 3.
Therefore, here we restrict the following analysis to just the first irradiation location.
During the course of the experiment, the material underwent a permanent volume expansion and lattice distortion. 
The distortion is apparent by the peak splitting of the (200) peak, see Figure~\ref{fig:3}b as compared to Figure~\ref{fig:3}a, which corresponds to a breaking of cubic symmetry via an elongation of the $c$-axis producing a body centered tetragonal (bct) cell.
Such a splitting was not observed in previous studies of YH$_{3}$ in this pressure range. 
Alongside the symmetry lowering, the volume per Y atom is also seen to increase in direct correlation to fluence incident on the sample.
This correlation becomes clear when volume change is plotted against fluence incident on the sample, as seen in~Figure\,\ref{fig:3}d.
The key difference is observed between the low fluence and high fluence runs. 
Directly after a low fluence run the diffraction peaks assigned to YH$_3$ show very little or no change in position, confirming a negligible change in volume after exposure to the x-rays.
This phenomena is not entirely due to high fluence on the sample since the material at 104\,GPa and 125\,GPa experienced not only higher total fluence in a given train, but higher fluence in a single pulse than the material at 153\,GPa, yet remained unchanged in between trains.
Therefore, the peak shifts shown in Figure~\ref{fig:3}c can be attributed to a sluggish transition induced by electronic/chemical disturbances from high fluence x-rays.

Although the predominant phase of YH$_x$ from around 20 to 160\,GPa is $Fm\overline{3}m$ YH$_3$,~\cite{peng_hydrogen_2017, troyan_anomalous_2021, snider_synthesis_2021, kong_superconductivity_2021, purans_local_2021} there is a possible tetragonal $I4/mmm$ phase of YH$_3$ predicted by crystal structure searching as low as 50\,GPa which has been seen to be slightly more enthalpically favorable than the cubic phase at pressures above 100\,GPa.~\cite{peng_hydrogen_2017,liu_high-pressure_2017} 
However, the tetragonal phase of YH$_3$ is predicted to be slightly denser than the cubic phase, rather than the observed increase in volume.~\cite{liu_high-pressure_2017}
On the other hand, YH$_4$ is expected to take on an $I4/mmm$ structure at these pressures with a larger theoretical volume, although it has been claimed that due to an unusually high thermal expansion for YH$_3$,~\cite{troyan_anomalous_2021} both it and YH$_4$ have very similar volumes at room temperature in this pressure range.~\cite{kong_superconductivity_2021}

The unusually high thermal expansion of YH$_{3}$ measured in \citet{kong_superconductivity_2021} may be explained by partial uptake of hydrogen, leading to non-stoichiometric hydride intermediates between YH$_{3}$ and YH$_{4}$.
To that extent, a model was built to evaluate the volume evolution and symmetry breaking through partial substitution of H into YH$_{3+x}$ (0 $\geq{}x\geq{}$ 1) to form YH$_4$.
The fcc YH$_3$ structure can be represented as a bct $I4/mmm$ structure with $c/a=\sqrt{2}$ and the Wyckoff label of the octahedral interstitial site changing from 4b to 2b.
In this representation it can be seen that the change from YH$_3$ to YH$_4$ is that the $2b$ site splits to form a $4e$ site.
Similarly, the difference between the YH$_2$ and YH$_3$ structures is occupation of the $2b$ site.
The partial occupancies at the $4e$ ($2b$) site were modelled within the framework of the Virtual Crystal Approximation (VCA) with Quantum Espresso.~\cite{villa-cortes_superconductivity_2022}
In this approach, the ionic potential is represented by the pseudopotential generated for a virtual atom with fractional nuclear charge, which provides a realistic representation of the average electronic charge density in the system as it happens in the real alloy.
Within the context of the primitive unit cell representation of $I4/mmm$ structure, this can be done one of two ways by having partial hydrogens on both of the 4e positions or one fully occupied position and the second partially occupied.

We found that the model with two partially occupied atoms produced a significantly larger volume compared to the cubic structure even for very small concentrations, $x$, although the two models come in to agreement around x\,=\,0.7.
The second model predicts a cubic structure when $x$ is close to zero, but it rapidly distorts into a tetragonal structure with $c/a\sim$1.8 by x\,=\,0.1.
At x\,=\,1 (ie. YH$_4$) the predicted $c/a$ is 1.9, a value similar to the previous prediction of $c/a$=1.886 at 150\,GPa.~\cite{troyan_anomalous_2021}
Fig~\ref{fig:3}e shows the volume changes as a consequence of stoichiometry change across YH$_{2}$H$^\text{2b}_{x}$ and YH$_{2}$H$^\text{4e}_{1+x}$, $0 \leq x \leq 1$, in the bct ($I4/mmm$) crystal structure.
When compared to the experimental data, one can see that the increases in volume seen in Figure~\ref{fig:3}e correspond fairly well with the partially occupied 4e site model.
The partial lowering of stoichiometry from YH$_3$ to YH$_2$ was not seen to be a good fit as the $c/a$ ratio remained $\sqrt{2}$ across the whole substitution range along with a mostly lowered volume.
The gradually increasing volume can thus be attributed to an incremental partial transformation from YH$_3$ to YH$_4$, which supports that making stoichiometric hydrides is a sluggish process that is highly dependent on total energy into the system, pressure and kinetics.

Another interesting phenomena is seen in YH$_3$ at 153\,GPa. 
There are marked differences during the course of low fluence and high fluence trains.
Note that here we define low fluence to be a pulse train that delivers pulses less than or equal to 46\,$\mu$J/pulse.
This is best illustrated in the trains shown in Figure~\ref{fig:4}a, where the most notable feature is a shift of the YH$_3$ reflections to higher or lower angles based on the fluence incident on the sample. 
The starting and ending diffraction patterns shown in blue in the stack plot are the first pulse of the first train and the first pulse of the next train, respectively, and are assumed to be "cold", with close to room temperature conditions. 
The "hot" pulses, shown in red, are diffraction patterns taken (sequentially with respect to the y-axis) during the rest of the train.
During low fluence trains, the diffraction peaks are observed to shift to lower 2$\theta$, as illustrated by the centroids of the peaks shown in Figure~\ref{fig:4}b left, indicating a volume expansion during the pulses. 
The unit cell volume then returns to its original size as confirmed by the first pulse of the next train and therefore the volume expansion is transient when exposed to low fluence pulses. 
During a high fluence train, diffraction peaks are observed to shift to higher 2$\theta$, as illustrated by the centroids of the peaks shown in Figure~\ref{fig:4}b right, indicating a reduction of the volume of the lattice.  
Decreased intensity and broadening of the peaks could suggest the sample is melting and that observed features are coming from cold sample at the tails of the beam. 
However, if this were the case, one would expect the center of the peak to remain stationary, not contract.
At 104 and 125\,GPa, during irradiation, regardless of fluence, peaks shifted to lower angles, suggesting thermal expansion of the material.
However, the peaks always returned to their initial positions by the start of the next train, as seen in Figure~S3.

As modeled for other systems, when a material is exposed to photon density below a certain threshold, energy is lost to electron--ion interactions and thermal processes can prevail.~\cite{medvedev_femtosecond_2019}
However, when the material is exposed to photon density above a certain threshold, a distribution of non-thermal electrons is produced and those drive changes in the system.
Albeit most non-thermal processes are modeled for one pulse experiments where the observation window does not include thermalization of the system; that is not to say these non-thermal processes can not dominate and be the main driver for a reaction.
Therefore, it could be that non-thermal effects dominate under high fluence conditions at 153\,GPa. 

Temperature determination from streaked optical pyrometry (SOP) provided limited results for most of the runs due low intensity.
This could be a symptom of the latent heat of formation preventing a high enough rise in temperature to be detected or the x-ray energies being close to the Y K-edge that fluorescence dominated the signal (this can be compensated for using a technique described in \cite{euxfel_community_dynamic_2023}).
The most reliable temperature determination was produced at 104\,GPa in which the sample was exposed to 79\,$\mu$J/pulse for 200 pulses; calculated temperatures ranged from 2500\,--\,5500\,K.
However, such high temperatures are attributed to the long pulse train.
Above 104\,GPa, in all cases, sufficient thermal emission was not measured on the SOP transformation.
This, coupled with the fact that none of the trains exceeded 15 pulses indicates that the temperatures were most likely below at least that (3000\,K) and the samples did not experience extreme temperatures.~\cite{euxfel_community_dynamic_2023}

\section{Conclusions}

In summary, $Pm\overline{3}n$ WP-YH$_x$ and $I4/mmm$ YH$_4$ were synthesized from $Fm\overline{3}m$ YH$_3$ and excess hydrogen at 125 and 153\,GPa using x-ray induced photochemistry at the EuXFEL.
This marks the first experimental realization of a Weaire-Phelan type yttrium hydride and the lowest pressures observed for the formation of $I4/mmm$ YH$_4$. 
These unprecedented syntheses can be attributed to the unique mechanisms afforded by driving reactions with intense x-rays; that would not be accessed under equilibrium thermodynamic conditions with conventaional techniques such as laser heating.
Interestingly, YH$_6$ which is a predicted and measured thermodynamically accessible compound around the experimental pressures is not observed here, again showing the unique non-equilibrium pathways afforded by x-ray induced photochemistry.
At 153\,GPa low fluence pulse trains cause no discernable change to the YH$_3$ while high fluence pulse trains lead to a permanently increased volume and lattice distortion which our calculations indicate is a sluggish transformation of YH$_{3+x}$ towards YH$_{4}$ with partial H uptake driven by each high fluence pulse train.
It is unclear if all of these behaviors are due to total fluence on the sample, or fluence per pulse.
Modeling an entire train remains challenging due to the computational expense, but modeling of pathways as a function of time over many pulses would be informative.
Studies have even started to deconvolute the complex interactions in multi-element systems, however due to the complexity of our system we find that such models cannot be used here.
However, the results do suggest that thermal states are populated by low fluence exposures, while additional non-thermal states are induced by high fluence exposures.
By using x-ray irradiation to excite the sample, chemistry was demonstrated that previously had required substantially higher pressures, thus the results here can help further elucidate the fundamental understanding of metal hydride synthesis.

\section{Data availability} 

The data that support the findings of this study are available from the corresponding author upon reasonable request as well as from \url{https://doi.org/10.22003/XFEL.EU-DATA-002855-00}.

\section{Acknowledgements}

This work supported by the U.S. Department of Energy, Office of Basic Energy Sciences under Award Number DE-SC0020303, as well as funded in part by the Gordon and Betty Moore Foundation's EPiQS Initiative, Grant GBMF10731. CPS and KVL were supported in part by the U.S. Department of Energy, Office of Basic Energy Sciences under Award Number DE-SC0023355. The authors are indebted to the HIBEF user consortium for the provision of instrumentation and staff that enabled this experiment.
We acknowledge European XFEL in Schenefeld, Germany, for provision of X-ray free-electron laser beamtime at Scientific Instrument HED (High Energy Density Science) under proposal \#2855 and would like to thank the staff for their assistance.
Some of this work was performed using resources provided by the Cambridge Service for Data Driven Discovery (CSD3) operated by the University of Cambridge Research Computing Service (www.csd3.cam.ac.uk), provided by Dell EMC and Intel using Tier-2 funding from the EPSRC (capital grant EP/T022159/1), and DiRAC funding from the STFC (www.dirac.ac.uk). 

\textit{For the purpose of open access, the authors have applied a Creative Commons Attribution (CC BY) licence to any Author Accepted Manuscript version arising from this submission.}

%

\section{Methods}

\subsection{YH$_3$ precursor preparation}

Samples of 99$\%$ purity yttrium foil were loaded in modified BX-90 style diamond anvil cells (DAC), inside a glove box with an Ar environment.
DACs were then sealed and reopened under vacuum to be gas loaded in a hydrogen environment at 3\,kbar.
Rhenium was used as the gasket material throughout.
Samples were compressed to target pressures between 100\,--\,160\,GPa and then left at room temperature for 4\,--\,6 weeks.
All samples fully converted to YH$_3$, which was confirmed by X-ray diffraction (XRD)~\cite{kume_high-pressure_2007, machida_long-period_2007, palasyuk_hexagonal_2005, machida_x-ray_2006,kong_superconductivity_2021,snider_synthesis_2021,troyan_anomalous_2021, purans_local_2021}.
Pressure was determined prior to the experiments from either the pressure-induced shift in Raman mode of diamond, or the position of the hydrogen vibron, or both.

\subsection{XFEL experiment}

Experiments were performed at the High Energy Density (HED) scientific instrument at the European X-ray Free-Electron Laser Facility (EuXFEL) in Schenefeld, Germany.~\cite{liermann_novel_2021, zastrau_high_2021}
Three different samples of YH$_3$, compressed to 104, 125, and 153\,GPa, were irradiated in several different locations across the sample.
Samples were both excited and probed with 18\,keV x-rays delivered in 20\,fs pulses with a 4.5\,MHz repetition rate. 
The beam was focused by compound refractive lenses (CRL) to a spot size of $10\times10\,\mu$m$^2$.
Fluence incident on the sample was determined by an intensity and position monitor (IPM) located downstream of the CRLs and attenuators, but upstream of the sample (note these numbers do not take absorption from the diamond into account).
For each sample, \% transmission and pulse number were incrementally increased starting as low as 0.6\% and 2 pulses.
Diffraction images were collected on the adaptive gain integrating pixel detector (AGIPD)~\cite{allahgholi_adaptive_2019}.
Temperatures were determined by optical pyrometry using a streak camera following protocols described elsewhere~\cite{euxfel_community_dynamic_2023}.
Further details on temperature determination and data analysis can be found in the Supplemental Material.

Samples were exposed to series of x-ray pulses via pulse trains, with varying x-ray fluence (0.6\,--\,100\% transmission).
In a given experiment, x-ray fluence was increased incrementally up to 100\% transmission or until diamond failure occurred.
For a given x-ray fluence, the number of pulses were increased until the sample was saturated (\textit{i.e.} no further change was observed), which typically occurred within 15 pulses.
Trains with higher pulse number (up to 200 pulses) were run, but with no further effect on the sample.
In between most high fluence exposures, a low fluence exposure (typically 6\% transmission which equals to, on average 15\,$\mu$J/pulse, 2 pulses) was used to probe the sample without exciting further reaction.
Due to the timing of the x-ray pulses and pump-probe nature of the experiment, diffraction patterns are representative of the state of the sample produced by the previous pulse (or in the case of the first XRD pattern in a train, that of the previous train).

The energy of an individual pulse was determined by using the signal from the diode, calibrated and corrected for the finite response time of the diode. 
Diffraction images were collected at the trough of each pulse on the AGIPD.~\cite{allahgholi_adaptive_2019} 
The detector was calibrated with both CeO$_2$ and Cr$_2$O$_3$. Diffraction images were integrated using Dioptas and analyzed with GSAS II \cite{prescher_dioptas_2015, toby_gsas-ii_2013}.

A streak camera collected intensity information continuously for the duration of each train at 450-850\,nm.
Only one train had sufficient data to determine temperature. The train consisted of 200 pulses at an average of 79\,$\mu$J/pulse whose data can be seen in Fig. S1 which includes the image of the streak camera and the plot of calculated temperatures.
However, due to technical difficulties, not all trains had streak camera data.
Sample fluorescence contributed to the signal; which could positively or negatively affect the temperature, details on temperature calculations can be found in \citet{euxfel_community_dynamic_2023}.

\subsection{Calculations}

\subsubsection{Ephemeral Data Derived Potentials}

Ephemeral Data Derived Potentials (EDDPs) can be used to accelerate crystal structure prediction. We trained a Y-H potential using the iterative approach described in reference~\cite{pickard_ephemeral_2022}.
In summary, this begins with 1,0000 single-point-energy calculations of randomly generated structures on which an EDDP is trained.
In each iteration, 100 local minima are found by random searching using the current EDDP, and single-point-energies are calculated for 10 `shaken' structures in the vicinity of each minimum.
At the end of each iteration, a new EDDP is trained. 

We used 5 iterations and local minima found at pressures randomly chosen between 50 and 150\,GPa.
The form of the potential is naturally cut-off at a distance of 5\,\AA, containing 5 polynomials, and a neural network containing a single layer with 5 nodes.
The structures in the training data contained between 1-4 Y and 1-20 H atoms with atomic separations of between 0.5 and 2.5\,\AA and volumes of 17-23\,\AA$^3${} per Y atom and 2-3.6\,\AA$^3${} per H atom.
We also included some known pure hydrogen structures, optimised and shaken around the local minima.

To perform searches, we looked for structures with between 2-8 Y and 4-80 H atoms with automatically-generated minimum separations volumes (\texttt{\#MINSEP=AUTO}).
This search generated 50,000 structures, each optimised at 100\,GPa using the EDDP. For each stoichiometry sampled, we then took all structures within 10\,meV/formula unit from the minimum and performed single-point energy calculations using DFT.
From this data, we calculate the convex hull and take all structures which remain within 10\,meV/formula unit of the convex hull to perform a full DFT geometry optimisation.
Typically, since the EDDP has already optimised the geometry to close to the minima, the DFT geometry optimisations do not require many steps.
Convex hulls at 100 and 150\,GPa can be seen in the main text; and the convex hull at 200\,GPa can be seen in Figure~S7.

For DFT calculations used in the training and post-search steps, we used a plane-wave cutoff of 600\,eV and a k-point spacing of  0.03\,$\times 2\pi\text{\AA}^{-1}$. 

\subsubsection{Virtual Crystal Approximation}

The structural optimizations for the virtual crystal approximation simulations were carried out employing the variable-cell relaxation method within density functional theory (DFT)~\cite{kohn_self-consistent_1965} as implemented in the Quantum Espresso suite.~\cite{giannozzi_quantum_2009}
Exchange and correlation contributions were taken into account in the generalized gradient approximation (GGA) by the Perdew–Burke–Ernzerhof (PBE) functional~\cite{perdew_generalized_1996} using optimized norm-conserving Vanderbilt pseudopotentials~\cite{hamann_optimized_2013} from the PseudoDojo Library.~\cite{van_setten_pseudodojo_2018}
The energy cut-off was 80\,Ry for the calculation of all stoichiometries, while a Monkhorst–Pack $24\times24\times24$ $k$-point mesh was employed for the Brillouin zone (BZ) integration, with Gaussian smearing of 0.02\,Ry.~\cite{fu_first-principles_1983} 

\end{document}


\title{Ultra-fast yttrium hydride chemistry at high pressures via non-equilibrium states induced by x-ray free electron laser}

\author{Emily~Siska}
 \email{These authors contributed equally}
 \affiliation{\NEXCL}
 \author{G.~Alexander~Smith}
 \email{These authors contributed equally}
 \affiliation{\NEXCL}
 \affiliation{\LasVegasChem}
 \author{Sergio Villa-Cortes}
    \affiliation{\NEXCL}
\author{Lewis J. Conway}
 \affiliation{\Cambridge}
 \affiliation{\CJPJapan}
 \author{Joshua Van Cleave}
 \affiliation{\NEXCL}
  \affiliation{\LasVegasPhys}
\author{Sylvain Petitgirard}
 \affiliation{\ETH}
\author{Valerio Cerantola}
 \affiliation{\Euro,}
 \affiliation{\italy}
   \author{Karen Appel}
 \affiliation{\Euro}
        \author{Carsten Baehtz}
 \affiliation{\Euro}
        \author{Victorien Bouffetier}
 \affiliation{\Euro}
        \author{Anand Dwiwedi}
 \affiliation{\Euro}
        \author{Sebastian G\"ode}
 \affiliation{\Euro}
     \author{Taisia Gorkhover}
 \affiliation{\Ham}
  \author{Zuzana Konopkova}
 \affiliation{\Euro}
     \author{Mohammad Hosseini}
 \affiliation{\Des}
 \author{Rachel J. Husband}
 \affiliation{\Euro}
   \author{Stephan Kuschel}
 \affiliation{\Ham}
      \author{Torsten Laurus}
 \affiliation{\Des}
        \author{Motoaki Nakatsutsumi}
 \affiliation{\Euro}
\author{Cornelius Strohm}
 \affiliation{\Euro}
        \author{Jolanta Sztuk-Dambietz}
 \affiliation{\Euro}
  \author{Ulf Zastrau}
 \affiliation{\Euro}
 \author{Dean Smith}
    \affiliation{\NEXCL}
 \author{Keith V. Lawler}
 \affiliation{\NEXCL}
 \author{Chris J. Pickard}
 \affiliation{\Cambridge}
 \affiliation{\CJPJapan}
\author{Craig P. Schwartz}
 \email{craig.schwartz@unlv.edu}
 \affiliation{\NEXCL}
\author{Ashkan Salamat}
 \email{ashkan.salamat@unlv.edu}
  \affiliation{\NEXCL}
 \affiliation{\LasVegasPhys}

\maketitle

\section{Experimental Details}



        \begin{figure} [h!]
    \centering
 \includegraphics[width=0.50\textwidth]{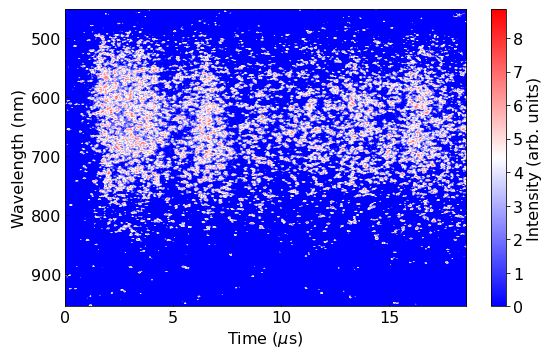}
 \includegraphics[width=0.90\textwidth]{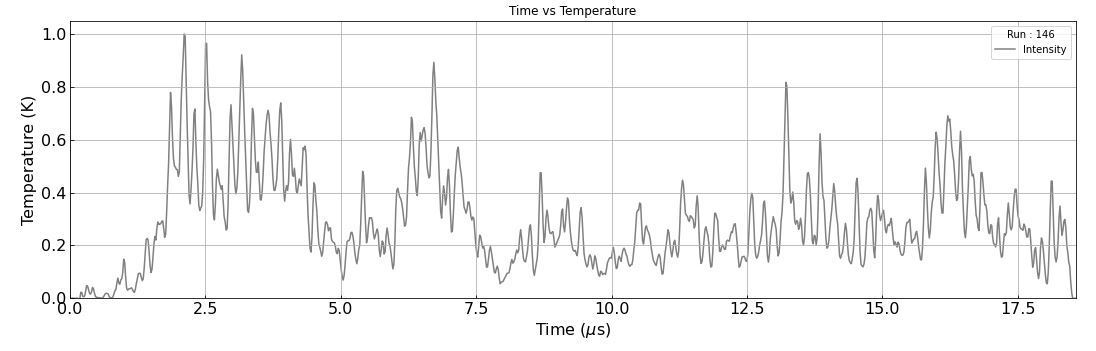}
    \caption{(a) Streak camera image and (b) plot of calculated temperatures \ES{Used as place holder until SOP analysis Notebook works}
    }
    \label{fig:Temp_data}
\end{figure}

\ES{Runs numbers of those featured in paper}




\section{Sample Details}
\subsection{104\,GPa}

\begin{figure} [h!]
    \centering
 \includegraphics[width=0.85\textwidth]{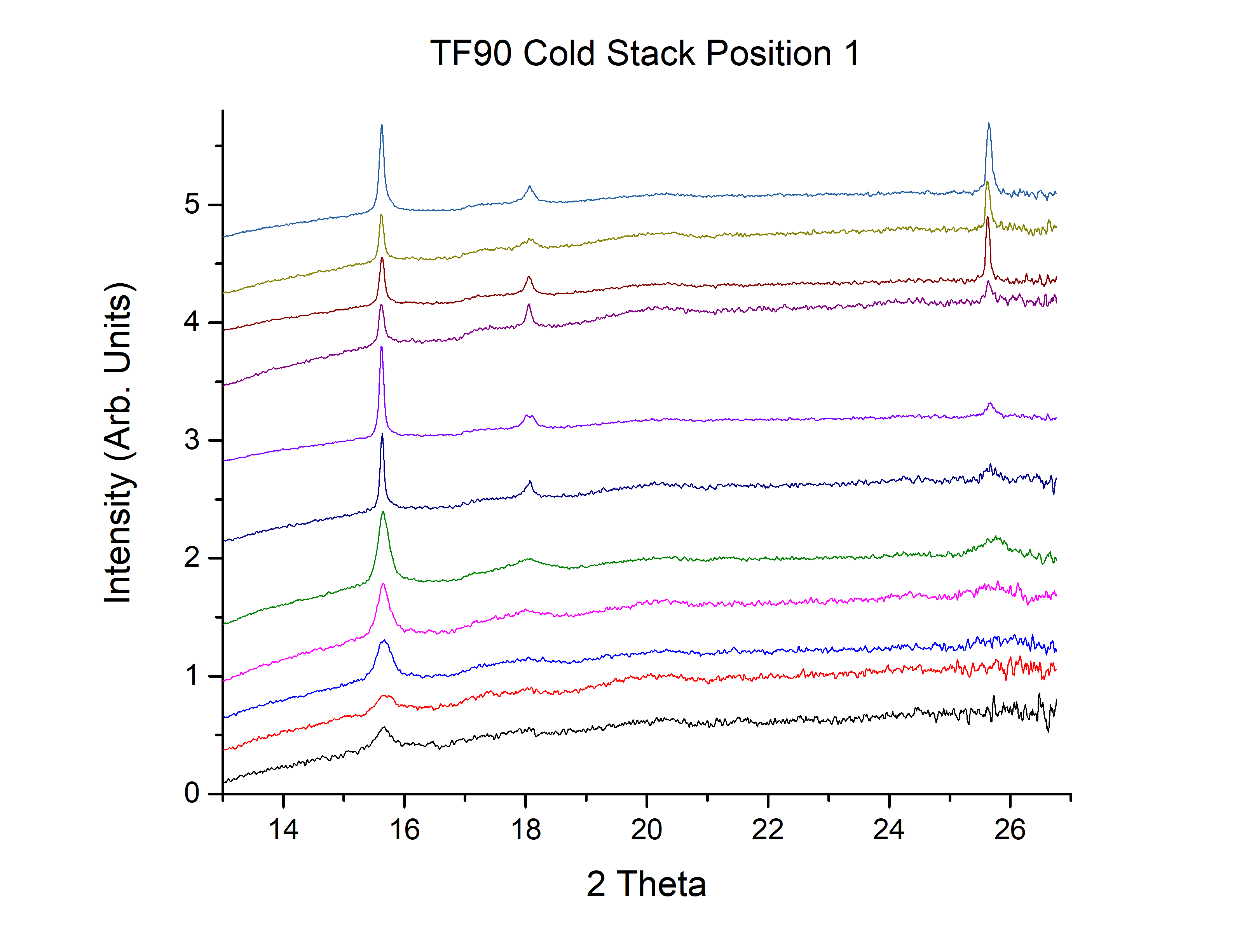}
  \includegraphics[width=0.85\textwidth]{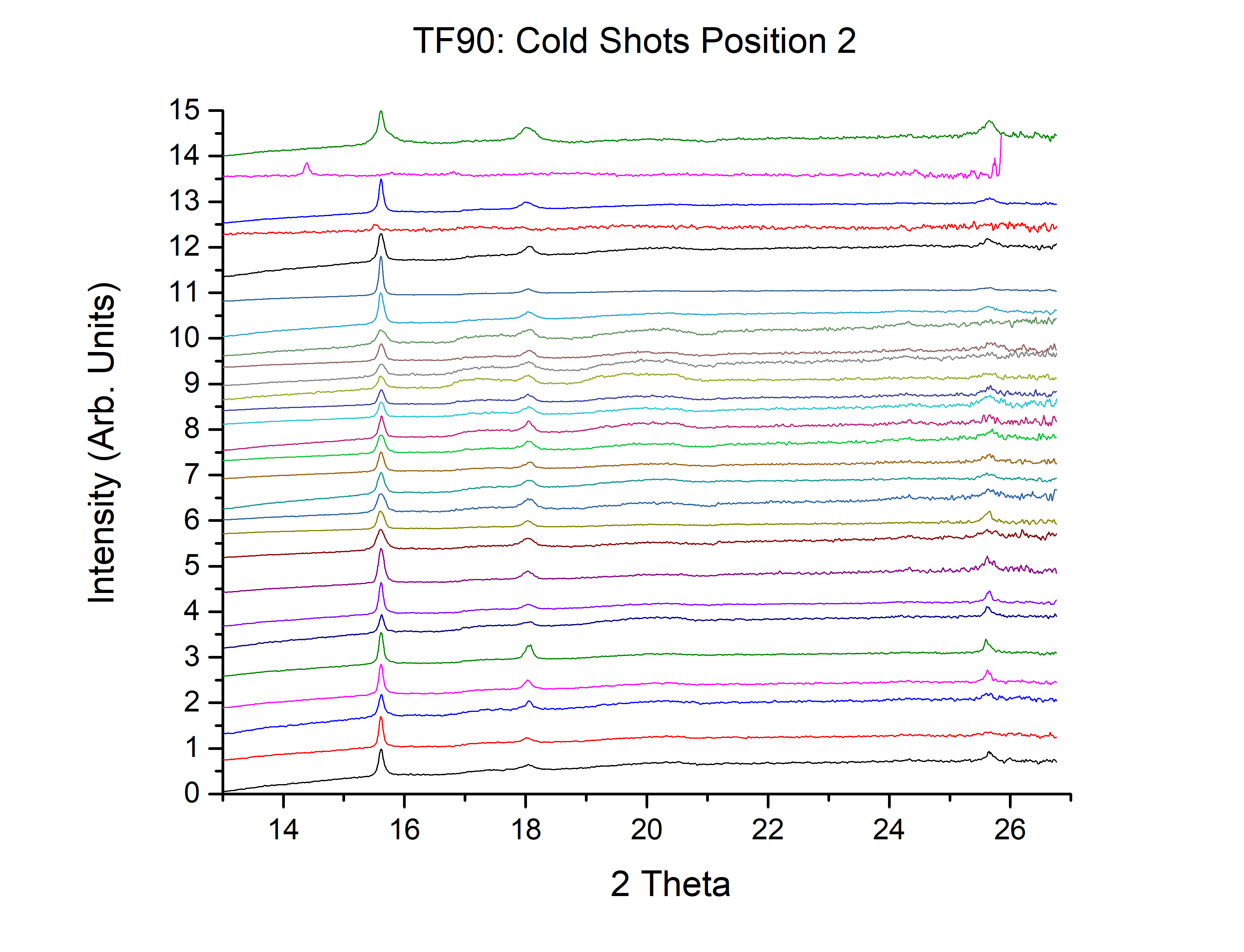}
    \caption{Consecutive XRD patterns of sample between each pulse train.
    }
    \label{fig:TF90 Pos1}
\end{figure}

Starting material $Fm\overline{3}m$ YH$_3$ at 104\,GPa was confirmed by XRD at EuXFEL. The material was exposed to 2 to 100\,\% transmission in 1 to 200 pulse trains at 2 different locations. Stack plots of the first XRD of each train can be seen in Fig.\,\ref{fig:TF90 Pos1}. In the stack plots, one can see the (111) peak of YH$_3$ in the initial XRD pattern, and the (200) and (220) peaks appear and persist throughout the experiment.

The sample, which had as much as 100\,\% transmission incident on the sample only had measurable streak camera data during a run at 40\% transmission, 200 pulses. 
Since most data was collected during trains below 70\% transmission, 15 pulses, we can conclude that the samples did not experience extreme temperatures (above 1500\,K) during the experiment. 
However, peaks did shift left during exposure, indicating expansion of the lattice - even during runs that had no measurable streak camera data, as seen in Fig\,\ref{fig:TF90 Fluence}. Showing high vs. low fluence trains illustrates the similar behavior of the material during trains, regardless of fluence incident on the sample. In both cases, the material undergoes thermal expansion like behavior. Although we do not rule out other processes that could be affecting the material.

        \begin{figure} [h!]
    \centering
 \includegraphics[width=0.6\textwidth]{Figures/High vs low fluence_104GPa.pdf}
    \caption{ XRD patterns taken during a single train of 15 pulses at (a) 35\,$\mu$J/pulse and (b) 152\,$\mu$J/pulse. The blue pattern at the bottom of the plot is the first XRD pattern take during the train and the blue pattern at the top of the plot is the first XRD pattern of the subsequent train - there to show the outcome of the train. The red XRD patterns are sequential XRD patterns in the train. 
    }
    \label{fig:TF90 Fluence}
\end{figure}

\subsection{125\,GPa}

\begin{figure} [h!]
    \centering
 \includegraphics[width=0.6\textwidth]{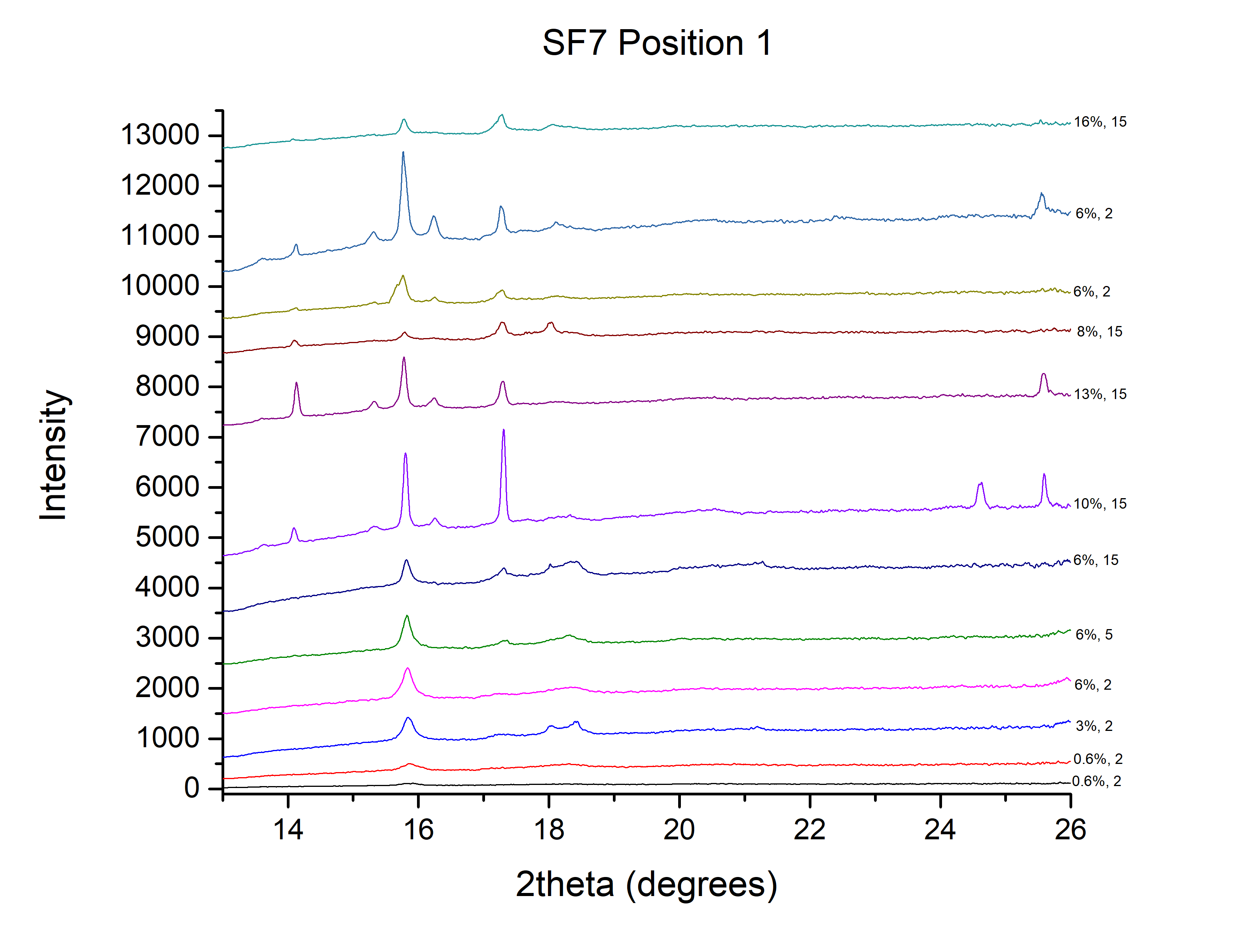}
  \includegraphics[width=0.6\textwidth]{Figures/SF7 Position 2 and 3 Cold Patterns.png}
  \includegraphics[width=0.6\textwidth]{Figures/SF7 Positions 4 7 Cold Patterns.png}
    \caption{"Unexcited" patterns (1$^{st}$ frame in a train). Beside each pattern is the \% transmission and number of shots from the previous run. Since the pattern is the 1$^{st}$ one in the train, it is a consequence on the previous exposure. 
    }
    \label{fig:SF7 Pos1}
\end{figure}

YH$_3$ at 125\,GPa was confirmed with XRD at the EuXFEL. The sample was exposed to 0.6 - 60 percent transmission with 2 - 55 pulse trains until diamond failure at 55\% transmission in a 15 pulse train. The sample was irradiated in 7 different locations, as seen in Fig.\,\ref{fig:SF7 Pos1}. 

All predicted compounds were evaluated against each new set of peaks that arose during the experiments. Two other structures fit observed peaks with acceptable volumes. The (111) reflections of both $F43m$ - YH$_9$ and $Fm\overline{3}m$ - YH$_3$ could be fit to observed features in XRD patterns at 125 and 153\,GPa. However, since only one peak as opposed to 6 (for the \textit{A}15 structure) could be fit; the WP-YH$_x$ structure was chosen over the other two. None of the clathrate structures, presented in \citet{peng_hydrogen_2017} were observed.

        \begin{figure} [h!]
    \centering
 \includegraphics[width=0.60\textwidth]{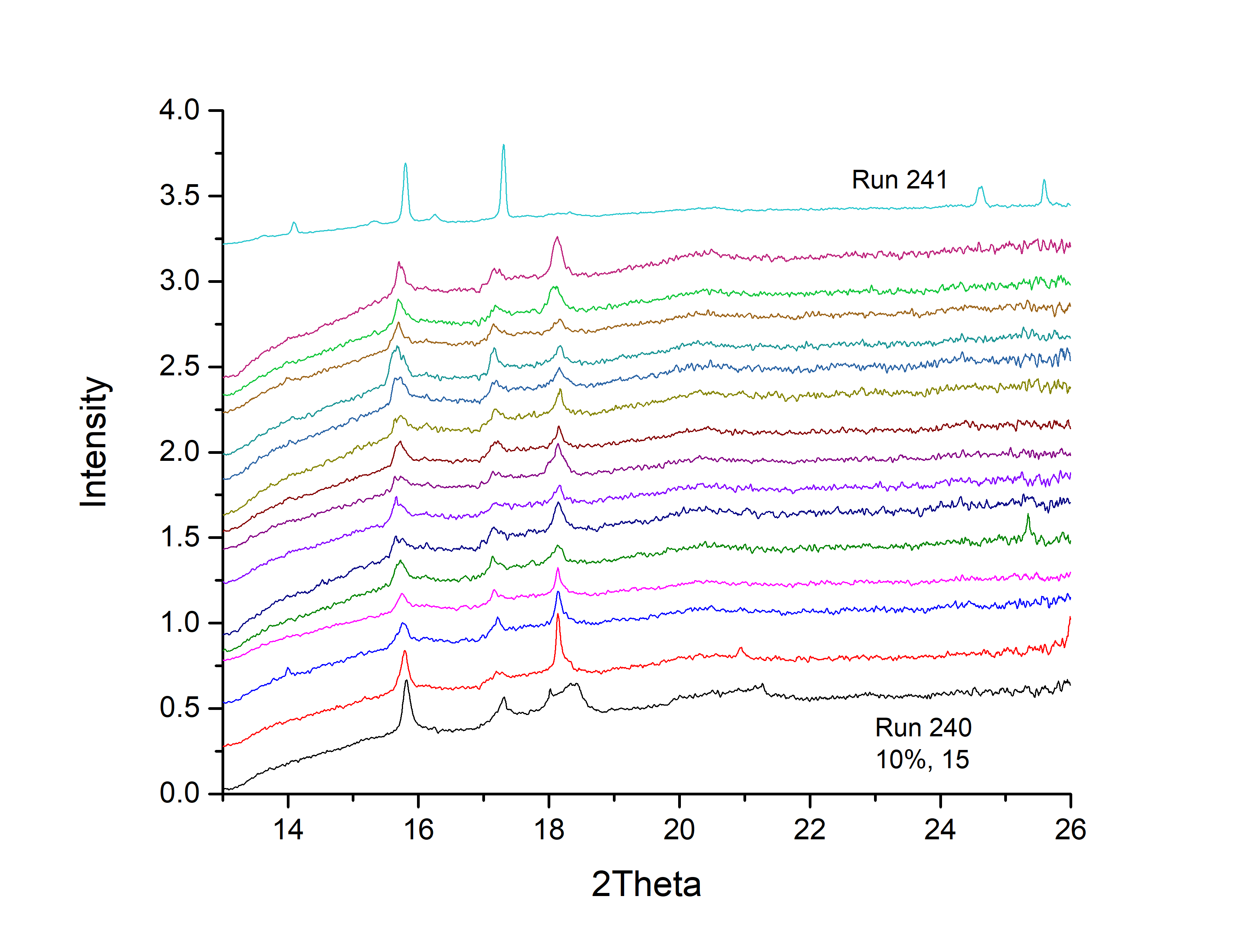}
    \caption{Patterns taken at the end of each pulse. The first pattern from run 241 shows the result of irradiation. 3 peaks attributed to YH$_3$ and the \textit{A}15 structure grow; vary in intensity with x-ray exposure. Also observed in other runs (e.i. 236 - 239). However, peaks loose significant intensity after irradiation. This is not the case for other peaks attributed to the \textit{A}15 structure, which exhibit nice, sharp peaks, but is lost during irradiation. This could be due to preferred orientation or the loss and gain of hydrogens in the material during and after irradiation
    }
    \label{fig:run 240}
\end{figure}
\FloatBarrier

\subsection{153\,GPa}

\begin{figure} [h!]
    \centering
 \includegraphics[width=0.60\textwidth]{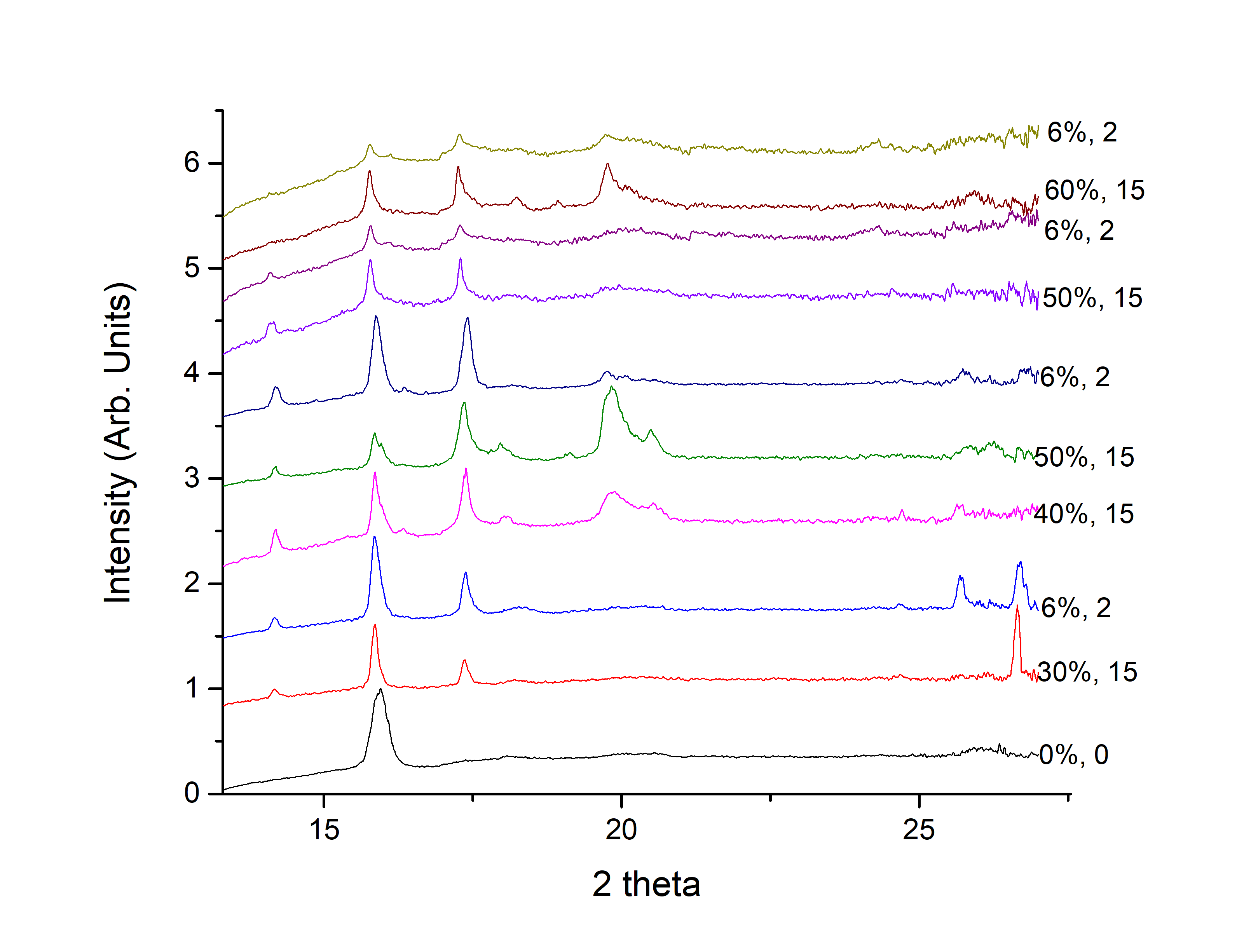}
    \caption{SF6, XRD pattern take between each consecutive pulse train.
    }
    \label{fig:SF6 location 3}
\end{figure}

YH$_3$ at 153\,GPa was confirmed with XRD at the EuXFEL.The sample was exposed to 6 - 70\,\% transmission over 2 - 25 pulses before diamond failure at 70\% transmission for 15 pulses. The sample was heated in 3 different locations, as seen in Fig.\,\ref{fig:SF6 location 3} A majority of the time, the system was saturated by 10 pulses, regardless of \% transmission.

The sample was irradiated in 3 different locations, with the first location having distinct results from the other 2. 

At location 1, the sample initially experienced 15\,$\mu$J/pulse, then fluence was increased slowly up to 126\,$\mu$J/pulse. 
After the sample was exposed to 10\,$\mu$J/pulse, the faint appearance of WP-YH$_x$ was observed. 
The phase did not grow in intensity with subsequent trains and the three high angle reflections were never observed.
This is in contrast to the first irradiated position in the 125\,GPa DAC, where WP-YH$_x$ is observed immediately, with one reflection observed during the second train at 4\,$\mu$J/pulse.
YH$_4$, along with the 3 other reflections for WP-YH$_x$ are observed slightly later when exposed to 28\,$\mu$J/pulse. 
YH$_4$, however, is never observed at 153\,GPa, nor is a complete pattern ever obtained for WP-YH$_x$.
In fact, the reflections assigned to WP-YH$_x$ are gone completely gone by train 18 (96\,$\mu$J/pulse) in the 25 train experiment.



\FloatBarrier

\begin{figure} [h!]
    \centering
 \includegraphics[width=0.90\textwidth]{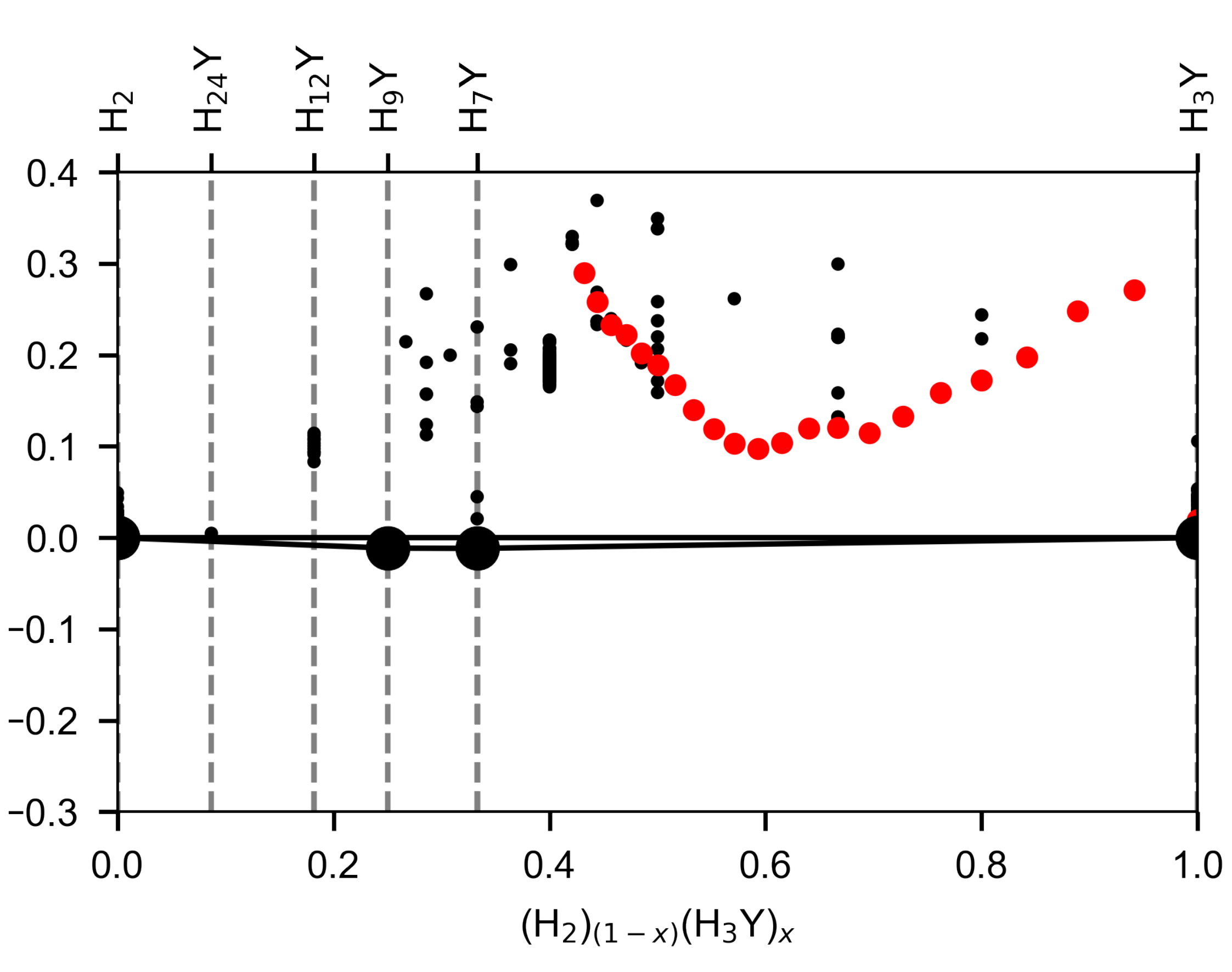}
    \caption{Convex hull of the (H$_{3}$Y)$_{(1-x)}$(H$_{2}$)$_{x}$ system at 200\,GPa, showing the metastability of WP-YH$_{x}$. Red circles show YH$_{x}$ species in which the Y atoms adopt an \textit{A}15-type lattice.
    }
    \label{fig:Convex_200}
\end{figure}

\FloatBarrier
\bibliography{references}